\documentclass[aps,prd,preprint,nofootinbib]{revtex4}
\usepackage{amsfonts}
\usepackage{mathrsfs}
\usepackage{graphicx}
\usepackage{amsmath}
\usepackage{amssymb}
\usepackage{epsfig}
\usepackage{graphicx}
\newcommand{\bqa}{\begin{eqnarray}}
\newcommand{\eqa}{\end{eqnarray}}
\newcommand{\beq}{\begin{equation}}
\newcommand{\eeq}{\end{equation}}
\graphicspath{{fig/}{dia/}} \DeclareGraphicsExtensions{.eps}
\usepackage{subfig}
\begin{document}
\baselineskip 20pt
\title{NLO QCD Corrections for $\chi_{cJ}$ Inclusive Production at \\ $B$ Factories\\}

\author{\vspace{1cm} Long-Bin Chen$^1$\footnote{
chenglogbin10@mails.ucas.ac.cn}, Jun Jiang$^1$\footnote{
jiangjun13b@mails.ucas.ac.cn} and Cong-Feng
Qiao$^{1,2}$\footnote{qiaocf@ucas.ac.cn, corresponding author} \\}

\affiliation{$^1$School of Physics, University of Chinese Academy of
Sciences,  Yuquan Road 19A, Beijing 100049, China}

\affiliation{$^2$CAS Center for Excellence in Particle Physics, Beijing 100049, China}

\begin{abstract}
{~~~\\[-3mm]}

The next-to-leading order (NLO) quantum chromodynamics (QCD) corrections for $\chi_{cJ}(^3P_J^{[1]},^3S_1^{[8]})$, the P-wave charmoniums inclusive production at $B$ factories are calculated utilizing the non-relativistic QCD (NRQCD) factorization formalism. Large NLO corrections are found, especially for $^3P_0^{[1]}$ and $^3S_1^{[8]}$ configurations.
Numerical evaluation indicates that the total cross sections of $\chi_{cJ}$ inclusive production processes are at the order of $10$fb, which are accessible in BELLE II(super-B) experiment.

\vspace {7mm} \noindent {PACS number(s): 13.66.Bc, 12.38.Bx,
14.40.Pq }

\end{abstract}
\maketitle

The advent of NRQCD factorization formalism placed the heavy quarkonium physics on a more solid ground \cite{NRQCD}. In the framework of NRQCD, nevertheless there are still many open questions about the quarkonium production and decay. A number of investigations indicate that the leading-order (LO) QCD calculations are inadequate to explain experimental
data. It seems so far that most of the discrepancies between LO
calculation and experimental observation can be rectified by including
higher order corrections, which has encouraged more NLO QCD calculations on quarkonium production and decay. On this point, one typical example is the double charmonium production at $B$ factories
\cite{bralee,hqiao,ZYJ,zyjchao,gbwang,GB1,qiao}.

The charmonium production at $B$ factories is one of the most
interesting and challenging problems in quarkonium physics. The
observed cross sections of charmonium production processes
$e^+e^-\rightarrow J/\psi+\eta_c$ and $e^+e^-\rightarrow
J/\psi+c+\bar{c}$ are much large than the LO QCD
theoretical results \cite{belle,babar,hechao,bralee,hqiao}. Through tedious investigations on the NLO QCD
corrections for these processes \cite{ZYJ,zyjchao,gbwang}, people found that the large gaps between theory and experiment can be greatly narrowed almost to non-existence. Though at the moment there have not been much data collected, the exclusive processes $e^+e^-\rightarrow J/\psi+\chi_{cJ}(J=0,1,2)$ at $B$ factories were investigated up to NLO accuracy \cite{chao,dfjia}. The large NLO QCD corrections to S-wave charmonium productions, exclusively and inclusively, enlighten us that higher order corrections to P-wave charmonium inclusive production must be very, if not more, important, not to mention here the color-octet contribution will be non-negligible. In the literature \cite{lhch}, the LO theoretical estimation for $e^+e^-\rightarrow \chi_{cJ}+c+\bar{c}(J=0,1,2)$ processes was made. For the aims of phenomenological study and the deeper understanding of NRQCD framework, in this work we calculate the NLO QCD corrections for the P-wave charmonium $\chi_{cJ}(^3P_{J}^{[1]}, ^3S_1^{[8]})$ inclusive production processes $e^+ e^- \rightarrow \gamma^*\rightarrow \chi_{cJ} + c\bar{c}(g g; q\bar{q})+X(J=0,1,2;\ q=u,d,s)$.

\begin{figure}[thbp]
\begin{center}
\includegraphics[scale=0.5]{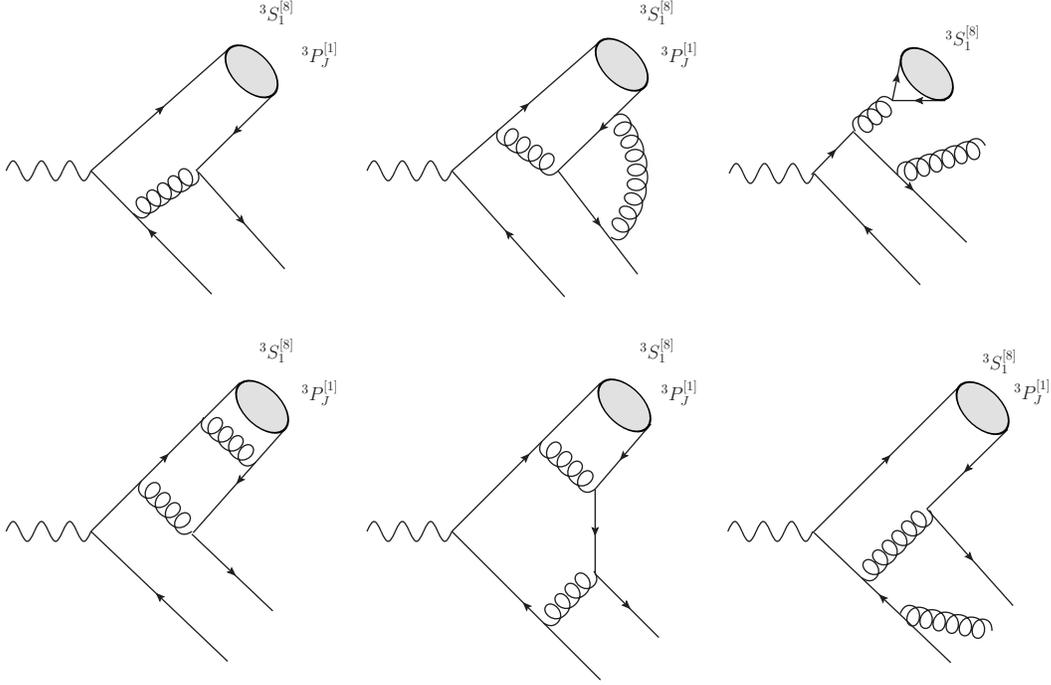}
\caption{Typical Feynman diagrams for $\gamma^*\rightarrow
\chi_{cJ}(^3P_{J}^{[1]},^3S_1^{[8]})+c\bar{c}+(g)$ . \label{fig1}}
\end{center}
\end{figure}

\begin{figure}[thbp]
\begin{center}
\includegraphics[scale=0.6]{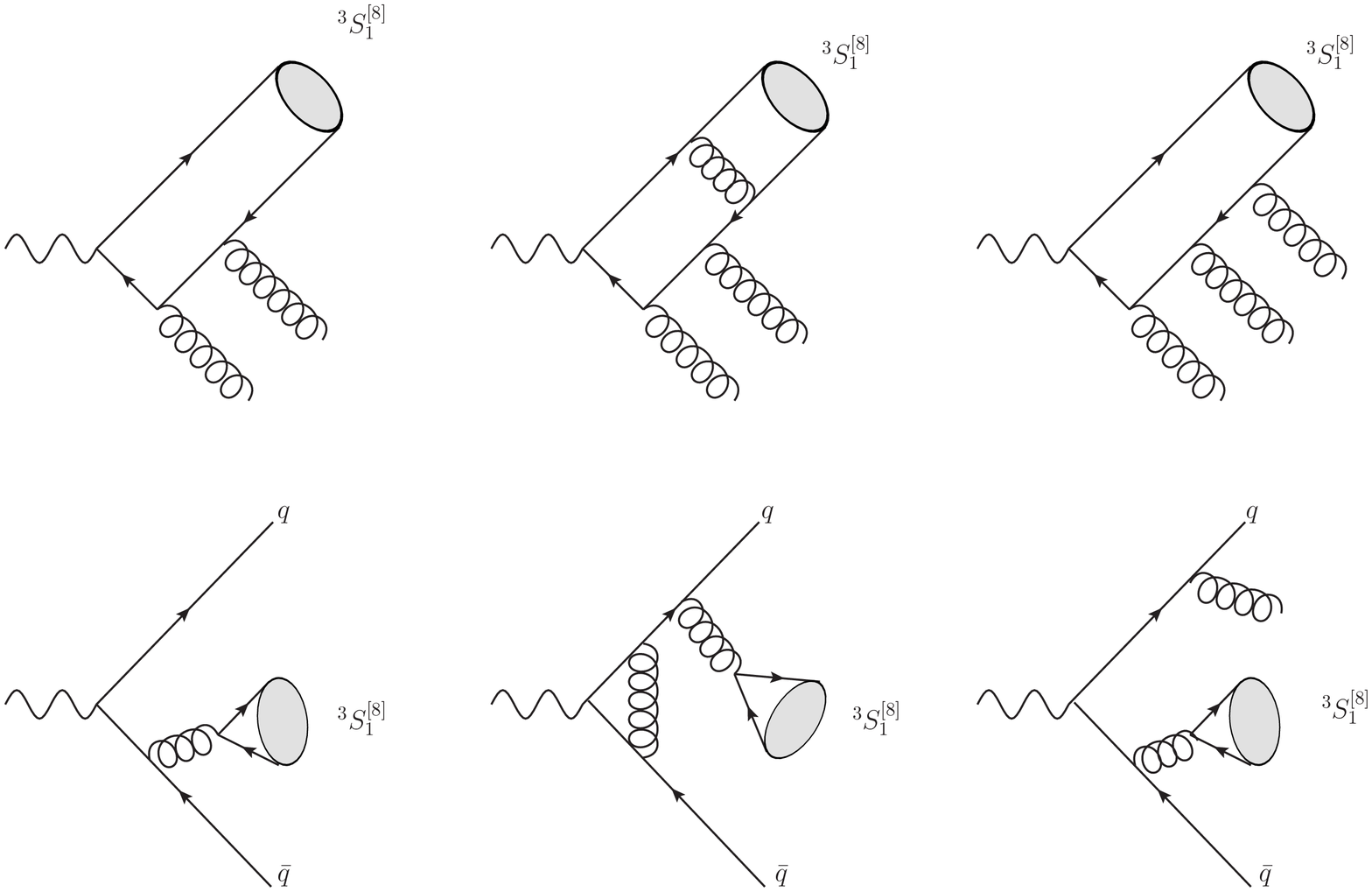}
\caption{Sample Feynman diagrams for $\gamma^*\rightarrow
\chi_{cJ}(^3S_1^{[8]}) + gg(q \bar{q}) + (g)$. \label{fig2}}
\end{center}
\end{figure}

In the calculation, \textbf{Mathematica} package of
\textbf{FeynArts} \cite{feynarts} was used to generate the LO and NLO
Feynman diagrams, as schematically shown in Fig.1 and Fig.2. The
standard form of quarkonium spin projection operator was adopted \cite{pchoaleb,bodwpe,bralee}. For color-singlet and spin-triplet
case of our concern, it reads:
\bqa
v(\bar{p})\bar{u}(p) =\frac{1}{4\sqrt{2}E(E+m)}(\not\!\bar{p}-m_c)
\not\!\epsilon_S^*(\not\!P+2E)(\not\!p+m_c)
\otimes\bigg(\frac{\bf{1}_c}{\sqrt{N_c}}\bigg)\ .
\label{projector}\eqa Here $p=\frac{P}{2}+q,~\bar{p}=\frac{P}{2}-q$
respectively are the momenta of quark and antiquark,
$\not\!\epsilon^*_S$ is a spin polarization vector,
$E^2=P^2/4=m_c^2-q^2$, $N_c=3$, and $\bf{1}_c$ represents the unit
color matrix. For color-singlet and spin-singlet state, the
projection operator may be obtained by replacing the $\not\!\epsilon_S^*$
in Eq.(\ref{projector}) by a $\gamma_5$, while for color-octet state
the color matrix $\bf{1}_c$ should be substituted by $\sqrt{2}\bf{T}_c$, the Gell-Mann matrices.

Employing the method used in Refs.\cite{bralee,pchoaleb} and by virtue of the symbolic computation package \textbf{FeynCalc} \cite{Feyncalc},
the Born level results are readily obtained. For NLO QCD
corrections, more complicated procedures have to be taken. The NLO corrections to the inclusive processes include two parts, the virtual and real corrections to the LO result. With virtual corrections, the cross section can be formulated as:
\bqa
{\rm d}\sigma_{virtual}
&=&\frac{1}{4}\ \frac{1}{2s} \ \sum \ 2 \ {\rm
Re}(\mathcal{M}_{Born}^*\mathcal{M}_{Virtual}) \rm{d}PS_3\ ,
\eqa
while with real corrections the cross section reads as:
\bqa
{\rm d}\sigma_{real} &=&\frac{1}{4}\ \frac{1}{2s}\ \sum
|\mathcal{M}_{real} |^2 \rm{d}PS_4\ .
\label{realcrosssection}
\eqa
Here $\rm{d}PS_3$ and $\rm{d}PS_4$ stand for three- and four-body phase spaces respectively. Note that for $n$ gluons production processes we should also multiply a factor of $\frac{1}{n!}$ in the phase space.

We use the phase space slicing method with two cutoffs in the calculation of real corrections \cite{twocut}. The outgoing gluon with energy $p_g^0<\delta$ is considered to be soft, while $p_g^0>\delta$
the gluon will be taken as a hard one. The $\delta$ here is a small quantity with energy-momentum unit. Under the soft condition of $p_g^0<\delta$ and in the Eikonal approximation,
\bqa \rm{d}PS_4|_{soft} = \rm{d}PS_3\frac{\rm{d}^3p_g}{(2\pi)^3p_g^0}|_{p_g^0<\delta}\ .
\label{reduction1}
\eqa
With the help of Eq.(\ref{reduction1}), the soft divergent, collinear divergent and finite parts in real correction Eq.(\ref{realcrosssection}) will be properly separated through the method outlined in Ref.\cite{twocut}.

Throughout our calculation, the self-developed codes based on \textbf{FeynCalc} \cite{Feyncalc} are used to trace the
matrices of spin and color, and to perform derivative on the
heavy quark relative momentum $q$ within quarkonium. The Mathematica
package \textbf{$\$$Apart} \cite{apart} reduces the propagators of
individual one-loop diagrams. After these procedures, the NLO virtual cross-section $\mathcal{M}_{Born}^*\mathcal{M}_{Virtual}$ is then expressed as linear combinations of one-loop integrals, i.e.
\bqa
I^N_0(D,\{n_i\})\equiv\int\frac{\rm{d}^Dq_1}{(2\pi)^D}\frac{1}{D_1^{n_1}D_2^{n_2}\cdots
D_N^{n_N}}\mid_{N\leq4}\ .
\eqa
The linearly independent propagators have the form
$D_i\equiv(q_1+r_i)^2-m_i^2$ and the index $n_i$ can be any
integers except 0. The package \textbf{Fire} \cite{fire} is then employed
to reduce all one-loop integrals $I^N_0(D,\{n_i\})$  to typical
master-integrals $(A_0,B_0,C_0,D_0)$, and the finite part of the master-integrals is computed by \textbf{LoopTools} \cite{looptools}.

We performed the calculation in the Feynman gauge, and the conventional dimensional regularization with $D = 4-2\epsilon$ was adopted in regularizing the ultraviolet and infrared divergences. In the end, all ultraviolet divergences are completely canceled by counter terms, while the Coulomb singularities are factorized out and attributed to the NRQCD long-distance matrix elements. Infrared divergences arising from loop integration, phase space and counter terms are partially canceled with each other. For $^3S_1^{[8]}$ the IR divergences totally canceled out in the calculation. After combining various infrared divergences, for color-singlet $^3P_J^{[1]}$ production processes one obtains divergent terms
proportional to the Born level amplitude of the color-octet $^3S_1^{[8]}$ process, and these divergent terms are then attributed to the
NLO color-octet NRQCD matrix elements $\langle0|\mathcal
{O}^{\chi_{cJ}}(^3S_1^{[8]})|0\rangle$ \cite{lmchao}. By this procedure, all the infrared divergences appearing in the NLO correction for $^3P_J^{[1]}$ and $^3S_1^{[8]}$ states production cancel out completely in the end.

The ultraviolet and infrared divergences exist also in the
renormalization constants $Z_2, Z_3, Z_m, Z_g$, corresponding
respectively to the quark field, gluon field, quark mass, and strong
coupling constant $\alpha_s$. Among them, the
$Z_g$ is defined in the modified-minimal-subtraction
$(\overline{MS})$ scheme, while the other three are in the on-shell (OS)
scheme. Thereafter, the counter terms read:
\begin{eqnarray}
\delta Z_2^{\rm OS}&=&-C_F\frac{\alpha_s}{4\pi}
\left[\frac{1}{\epsilon_{\rm UV}}+\frac{2}{\epsilon_{\rm IR}}
-3\gamma_E+3\ln\frac{4\pi\mu^2}{m^2}+4\right],
\nonumber\\
 \delta Z_m^{\rm OS}&=&-3C_F\frac{\alpha_s}{4\pi}
\left[\frac{1}{\epsilon_{\rm
UV}}-\gamma_E+\ln\frac{4\pi\mu^2}{m^2} +\frac{4}{3}\right], \nonumber\\
 \delta
Z_3^{\mathrm{OS}}&=&\dfrac{\alpha_s}{4\pi}\biggl[(\beta'_0-2C_A)(\dfrac{1}{\epsilon_{UV}}
-\dfrac{1}{\epsilon_{IR}}) -\dfrac{4}{3}T_F(\dfrac{1}{\epsilon_{UV}} -\gamma_E +\ln\dfrac{4\pi \mu^2}{m_c^2}) \biggr],
 \nonumber\\
  \delta Z_g^{\overline{\rm MS}}&=&-\frac{\beta_0}{2}\,
  \frac{\alpha_s}{4\pi}
  \left[\frac{1}{\epsilon_{\rm UV}} -\gamma_E + \ln(4\pi)
  \right].
\end{eqnarray}
The definitions of $\beta'_0$ and $\beta_0$ will be given afterwards.

We had taken several measures to check our calculation. We compared our LO result with Ref.\cite{hechao} and got an agreement while having the same inputs. We also made use of Helac-Onia \cite{helaconia} to calculations of LO processes and the hard parts of real corrections. We took several different values of the soft cut $\delta$ and
found the results are insensitive to the change. We apply our codes to the NLO calculations of $J/\psi\rightarrow e^+e^-$ and $\eta_c\rightarrow \gamma\gamma$ processes, and find our analytic results agree with those in the literature. Moreover, we found the NLO results for $\chi_{cJ}$ inclusive production processes can also be expressed in the form
\begin{equation}
\sigma_{NLO}= \sigma_{LO}(1 + \frac{\alpha_s}{\pi} (\frac{\beta_0}{2}\ln(\frac{\mu^2}{s})+C)),
\label{nlo-form}
\end{equation}
agree with the NLO result of $e^+e^- \rightarrow J/\psi+c+\bar{c}+X$ process \cite{zyjchao}.

To perform the numerical calculation, one needs first to fix the
inputs. In our numerical evaluation, the fine structure constant
$\alpha={1}/{137}$; the charm quark mass $m_c=1.5\pm0.1$ GeV; the
NRQCD matrix elements $\langle\mathcal {O}_1\rangle_{\chi_{cJ}}$ are
extracted from the $\chi_{c2}$ to two photon decay process, i.e.
$\Gamma(\chi_{c2}\rightarrow
\gamma\gamma)=\frac{128\pi\alpha^2\langle\mathcal
{O}_1\rangle_{\chi_{c2}}}{405m_c^4}(1-\frac{16\alpha_s}{3\pi})$
\cite{chao}, and it therefore is $\langle\mathcal
{O}_1\rangle_{\chi_{c2}}\equiv \langle\chi_{c2}|\mathcal
{O}^{\chi_{c2}}(^3P_2^{[1]})|\chi_{c2}\rangle=0.0166m_c^4~\text{GeV}$;
the magnitude of color-octet  matrix element $\langle0|\mathcal
{O}^{\chi_{c0}}(^3S_1^{[8]})|0\rangle=0.000748m_c^2~ \text{GeV}$
\cite{mawch}, which was obtained by fitting the $\chi_{cJ}$
hadroproduction theoretical result to the Tevetron data. The two-loop expression for the running coupling constant $\alpha_s(\mu)$ reads
\beq \frac{\alpha_s(\mu)}{4\pi}=\frac{1}{\beta_0 L}-\frac{\beta_1\ln
L}{\beta_0^3L^2}\ . \eeq Here, $L=\ln(\mu^2/\Lambda_{QCD}^2)$,
$\beta_0=(11/3)C_A-(4/3)T_Fn_f$, $\beta'_0=(11/3)C_A-(4/3)T_F(n_f-1)$, and
$\beta_1=(34/3)C_A^2-4C_FT_Fn_f-(20/3)C_AT_Fn_f$, with
$\Lambda_{QCD}$ to be $296$MeV \cite{PDG} and $n_f=4$, the number
of active flavors.

After substituting the above input parameters into the analytical
expressions for $\chi_{cJ}$ inclusive production and integrating over the phase space, the numerical results can be obtained. The cross sections for
$e^+e^-\rightarrow\gamma^*\rightarrow\chi_{cJ}(^3P_J^{[1]})+c+\bar{c}+X$
are presented in Table~I, where the uncertainty comes mainly from $m_c$. We
chose the renormalization scale $\mu$ running from $2m_c$ to
$\sqrt{s}/2$. For color-octet, since the matrix elements have the relation:
$\langle0|\mathcal {O}^{\chi_{c0}}(^3S_1^{[8]})|0\rangle :
\langle0|\mathcal {O}^{\chi_{c1}}(^3S_1^{[8]})|0\rangle :
\langle0|\mathcal {O}^{\chi_{c2}}(^3S_1^{[8]})|0\rangle = 1:3:5$, we merely
need to calculate one of the three processes. The cross sections for
$e^+e^-\rightarrow\gamma^*\rightarrow\chi_{cJ}(^3S_1^{[8]}) + c\bar{c}(g g; q\bar{q}) + X$ are presented in Table~II. The cross sections of $e^+e^-\rightarrow\gamma^*\rightarrow\chi_{cJ}(^3S_1^{[8]}) + q\bar{q} + X$ consist of the contributions from three flavors of light quarks $(u,d,s)$.
\begin{table}[h]
\caption{Cross sections of
$e^+e^-\rightarrow\chi_{cJ}(^3P_J^{[1]})+c+\bar{c}+X$ at leading
order and next-to-leading order, all in units of fb, with
$m_c=1.5\pm 0.1$GeV, and $\mu \in \{2m_c,\sqrt{s}/2 \}$.}
\begin{center}
\renewcommand\arraystretch{1.4}
  \begin{tabular}{|l|c|c|c|c|c|c|}
    \toprule
    $\ \ \sigma$(fb)& LO$\chi_{c0}^{[1]}$ & NLO$\chi_{c0}^{[1]}$ & LO$\chi_{c1}^{[1]}$
    & NLO$\chi_{c1}^{[1]}$ & $\textrm{LO} \chi_{c2}^{[1]}$ & NLO$\chi_{c2}^{[1]}$\\
    \hline
    $\mu = 2m_c$& $34.2^{+4.9}_{-4.6}$ &
    $84.3^{+20.7}_{-12.6}$ &
    $9.5^{+3.0}_{-2.5}$ & $4.0^{+0.4}_{-0.7}$
    &
    $4.4^{+1.3}_{-1.0}$ & $0.8^{+1.8}_{-0.8}$\\
    $\mu = \sqrt{s}/2$& $23.2^{+3.3}_{-3.1}$
    & $58.3^{+15.5}_{-7.3}$ & $6.4^{+2.1}_{-1.6}$
    & $5.3^{+1.3}_{-1.1}$ & $3.0^{+0.8}_{-0.7}$ & $1.9^{+0.6}_{-0.8}$\\
    \botrule
  \end{tabular}
\end{center}
\label{tab:cro1}
\end{table}
\begin{table}[h]
\caption{Cross sections of
$e^+e^-\rightarrow\chi_{cJ}(^3S_1^{[8]})+ c\bar{c}(g g; q\bar{q}) + X$ at leading order and next-to-leading order, all in units of fb, with
$m_c=1.5\pm 0.1$GeV, and $\mu \in \{2m_c,\sqrt{s}/2 \}$.}
\begin{center}
\renewcommand\arraystretch{1.4}
 $e^+e^-\rightarrow\chi_{cJ}(^3S_1^{[8]})+c+\bar{c}+X$
  \begin{tabular}{|l|c|c|c|c|c|c|}
    \toprule
    $\ \ \sigma$(fb)& LO$\chi_{c0}^{[8]}$ & NLO$\chi_{c0}^{[8]}$ & LO$\chi_{c1}^{[8]}$
    & NLO$\chi_{c1}^{[8]}$ & $\textrm{LO} \chi_{c2}^{[8]}$ & NLO$\chi_{c2}^{[8]}$\\
    \hline
    $\mu = 2m_c$& $0.60^{+0.25}_{-0.18}$ &
    $1.47^{+0.55}_{-0.59}$ &
    $1.8^{+0.75}_{-0.54}$ & $4.41^{+1.65}_{-1.77}$
    &
    $3.00^{+1.25}_{-0.90}$ & $7.35^{+2.75}_{-2.95}$\\
    $\mu = \sqrt{s}/2$& $0.41^{+0.17}_{-0.13}$
    & $1.02^{+0.39}_{-0.39}$ & $1.23^{+0.51}_{-0.39}$
    & $3.06^{+1.17}_{-1.17}$ & $2.05^{+0.85}_{-0.65}$ & $5.1^{+1.95}_{-1.95}$\\
    \botrule
  \end{tabular}
   $e^+e^-\rightarrow\chi_{cJ}(^3S_1^{[8]})+g+g+X$
  \begin{tabular}{|l|c|c|c|c|c|c|}
    \toprule
    $\ \ \sigma$(fb)& LO$\chi_{c0}^{[8]}$ & NLO$\chi_{c0}^{[8]}$ & LO$\chi_{c1}^{[8]}$
    & NLO$\chi_{c1}^{[8]}$ & $\textrm{LO} \chi_{c2}^{[8]}$ & NLO$\chi_{c2}^{[8]}$\\
    \hline
    $\mu = 2m_c$& $0.65^{+0.02}_{-0.03}$ & $1.31^{+0.06}_{-0.06}$
     &
    $1.95^{+0.06}_{-0.09}$ & $3.93^{+0.18}_{-0.18}$
    &
    $3.25^{+0.10}_{-0.15}$ & $6.55^{+0.30}_{-0.30}$\\
    $\mu = \sqrt{s}/2$& $0.44^{+0.02}_{-0.02}$
    & $0.91^{+0.05}_{-0.04}$ & $1.32^{+0.06}_{-0.06}$
    & $2.73^{+0.15}_{-0.12}$ & $2.20^{+0.10}_{-0.10}$ & $4.55^{+0.25}_{-0.20}$\\
    \botrule
  \end{tabular}
    $e^+e^-\rightarrow\chi_{cJ}(^3S_1^{[8]})+q+\bar{q}+X$
  \begin{tabular}{|l|c|c|c|c|c|c|}
    \toprule
    $\ \ \sigma$(fb)& LO$\chi_{c0}^{[8]}$ & NLO$\chi_{c0}^{[8]}$ & LO$\chi_{c1}^{[8]}$
    & NLO$\chi_{c1}^{[8]}$ & $\textrm{LO} \chi_{c2}^{[8]}$ & NLO$\chi_{c2}^{[8]}$\\
    \hline
    $\mu = 2m_c$& $1.82^{+0.52}_{-0.40}$ & $4.09^{+1.11}_{-0.85}$
     &
    $5.46^{+1.56}_{-1.20}$ & $12.27^{+3.33}_{-2.55}$
    &
    $9.10^{+2.60}_{-2.00}$ & $20.45^{+5.55}_{-4.25}$\\
    $\mu = \sqrt{s}/2$& $1.24^{+0.35}_{-0.27}$
    & $2.88^{+0.79}_{-0.60}$ & $3.72^{+1.05}_{-0.81}$
    & $8.64^{+2.37}_{-1.80}$ & $6.20^{+1.75}_{-1.35}$ & $14.4^{+3.95}_{-3.00}$\\
    \botrule
\end{tabular}
\end{center}
\label{tab:cro2}
\end{table}

From Tables~I to II we notice that after NLO QCD corrections,
cross sections for $^3P_0^{[1]}$ and $^3S_1^{[8]}$ production
are significantly enhanced, while cross sections for
$^3P_1^{[1]}$ and $^3P_2^{[1]}$ production are depressed. Relatively, the NLO corrections for $^3P_0^{[1]}$, $^3P_2^{[1]}$
and $^3S_1^{[8]}$ states are large, similar to cases of other charmonium production processes \cite{ZYJ,zyjchao,octet-bf}. In case we express the NLO cross sections as
$\sigma_{NLO}=\sigma_{LO} (1+\frac{\alpha_s}{\pi} (\frac{\beta_0}{2}\ln(\frac{\mu^2}{s})+C))$,
large $C$s yield from $^3P_0^{[1]}$ and $^3S_1^{[8]}$
production processes. Note that the large C not only produces notable NLO correction, but also induces evident renormalization scale dependence.
From Tables I to II one can also read that for
$\chi_{c0}$ production, the dominant
contribution comes from the color-singlet configuration, while for
$\chi_{c1}$ and $\chi_{c2}$ production the color-octet contributes
more. After adding all the channels of this work, the NLO inclusive production cross sections at $\mu=3\text{GeV}$ for $\chi_{cJ}(J=0,1,2)$ are $91.17$fb, $24.61$fb and $35.15$fb, respectively. The BaBar Collaboration \cite{ex-chic} had once searched for the prompt $\chi_{c1}$ and $\chi_{c2}$ production, and found no
significant signal; whereas they at $90\%$ confidence level presented the improved upper limits for the production of these two states, i.e. 77 fb for $\chi_{c1}$ and 79 fb for $\chi_{c2}$, which include our results. It is worth noting that the radiative processes $e^+ + e^- \rightarrow \chi_{cJ}+\gamma$ also contribute remarkably to the $\chi_{cJ}$ inclusive production at $B$ factories \cite{chyq}. However except for $\chi_{c1}$ production whose cross section is $10.9$fb, the cross sections for  $\chi_{c0}$ and $\chi_{c2}$ are much less in comparison with the processes of our concern.

\begin{figure*}[t]\centering
\subfloat[$\chi_{c0}( ^3P_0^{[1]})+ c+\bar{c}+ X$]{
  \begin{minipage}[t]{.5\linewidth}
    \includegraphics[width=\linewidth]{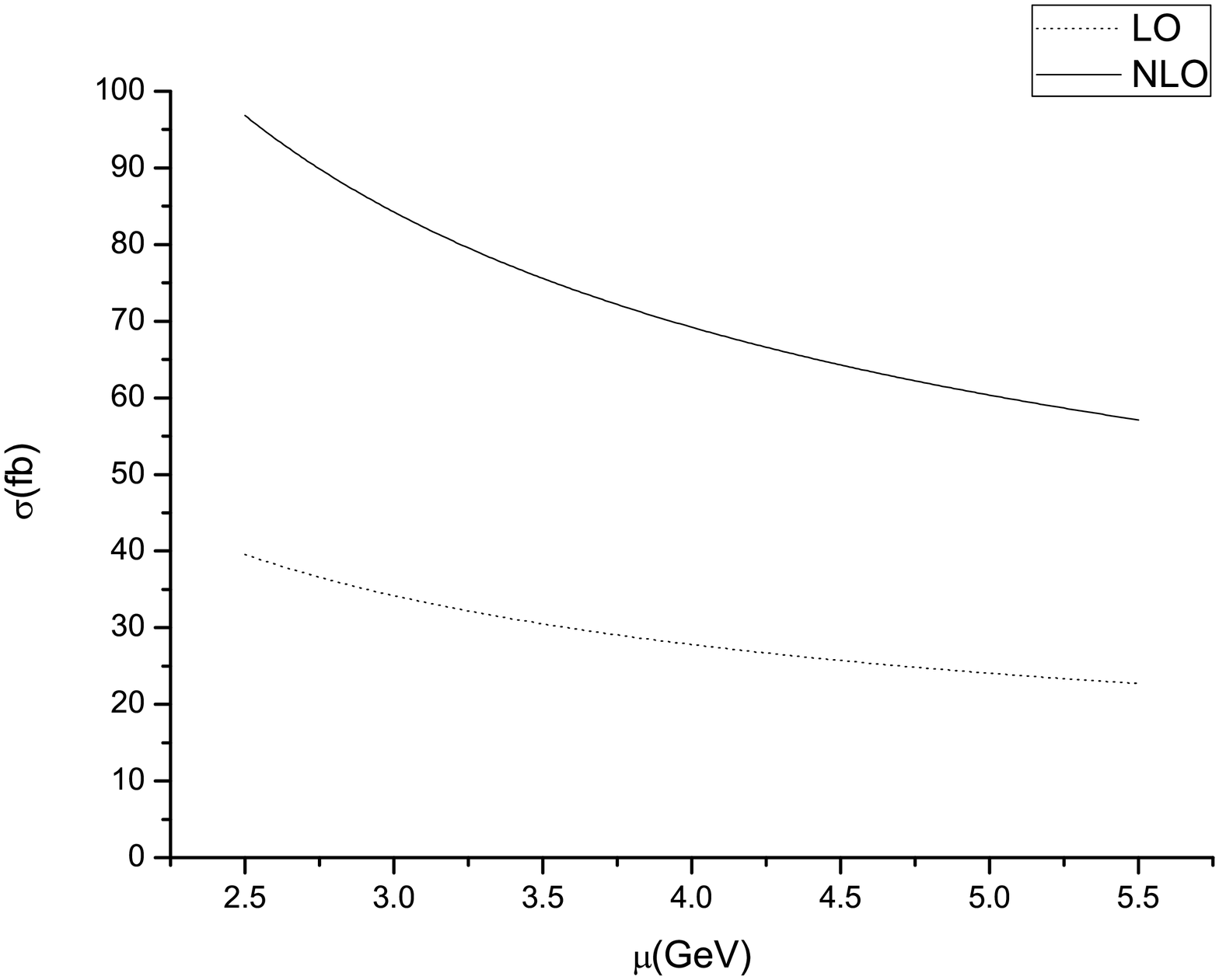}
  \end{minipage}
} \subfloat[$\chi_{c1}( ^3P_1^{[1]})+ c+\bar{c}+ X$]{
  \begin{minipage}[t]{.5\linewidth}
    \includegraphics[width=\linewidth]{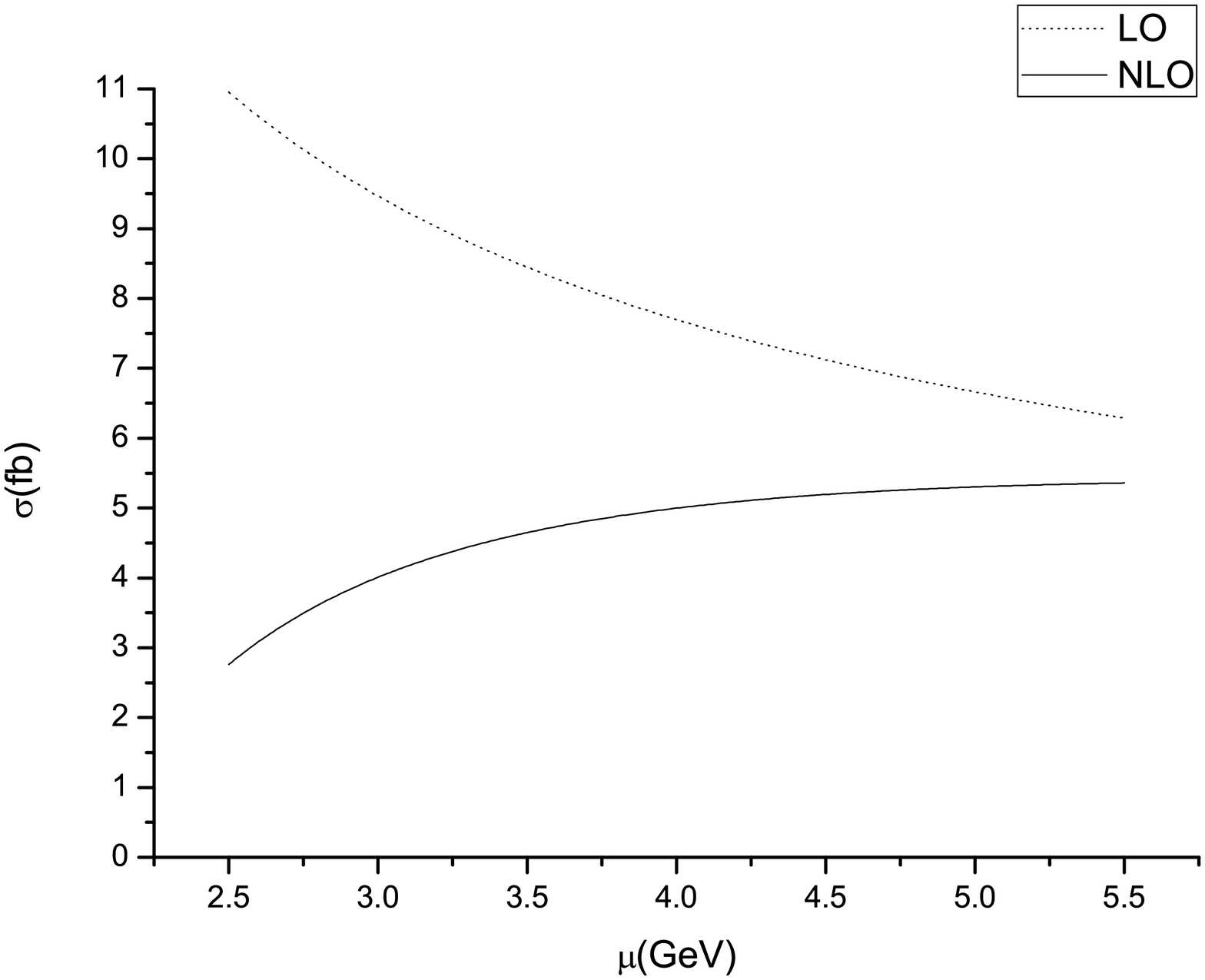}
  \end{minipage}
}\\
 \subfloat[$\chi_{c2}(^3P_2^{[1]})+ c+\bar{c}+ X$]{
  \begin{minipage}[t]{.5\linewidth}
    \includegraphics[width=\linewidth]{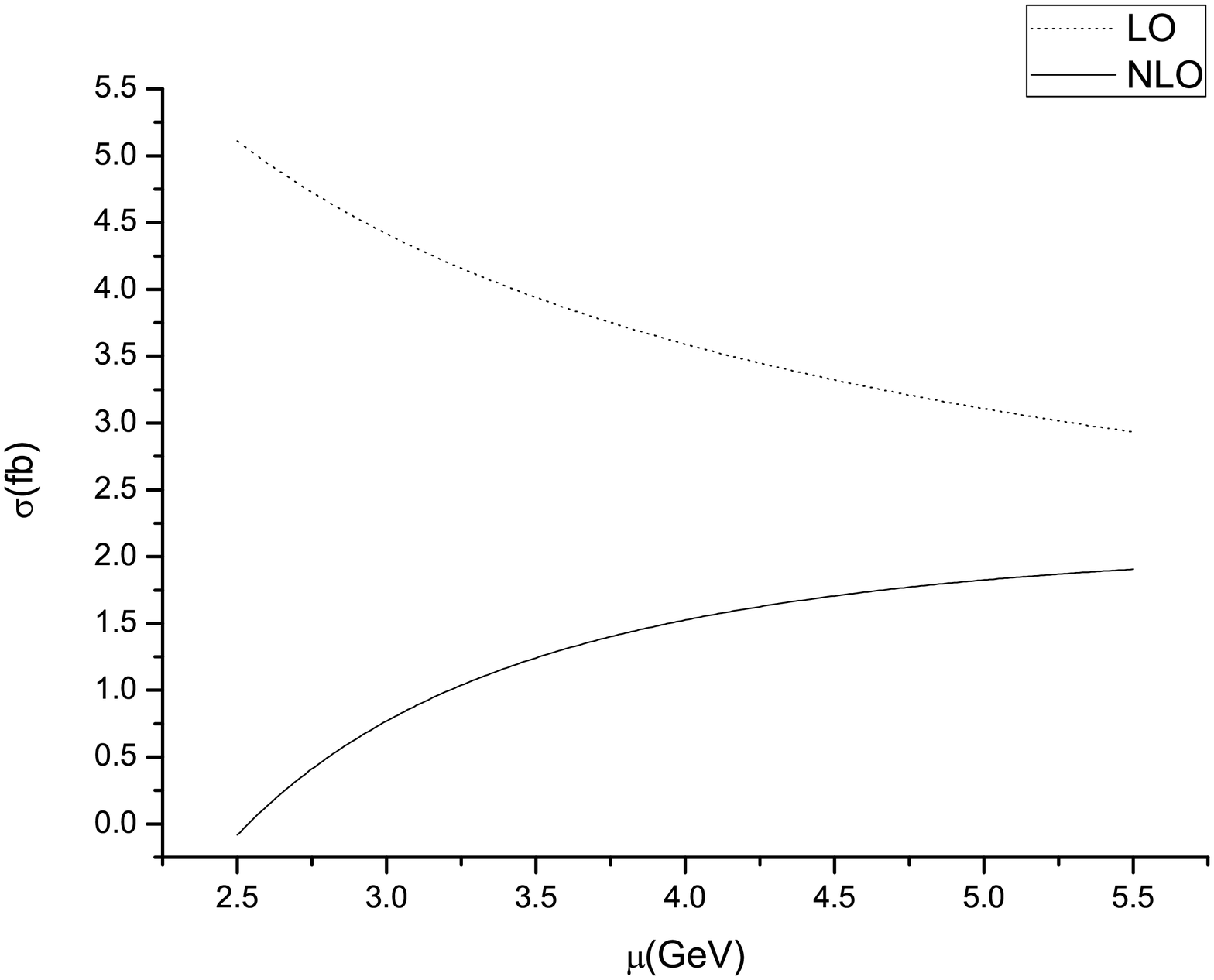}
  \end{minipage}
}
 \subfloat[$\chi_{c0}(^3S_1^{[8]})+ c+\bar{c}+ X$]{
  \begin{minipage}[t]{.5\linewidth}
    \includegraphics[width=\linewidth]{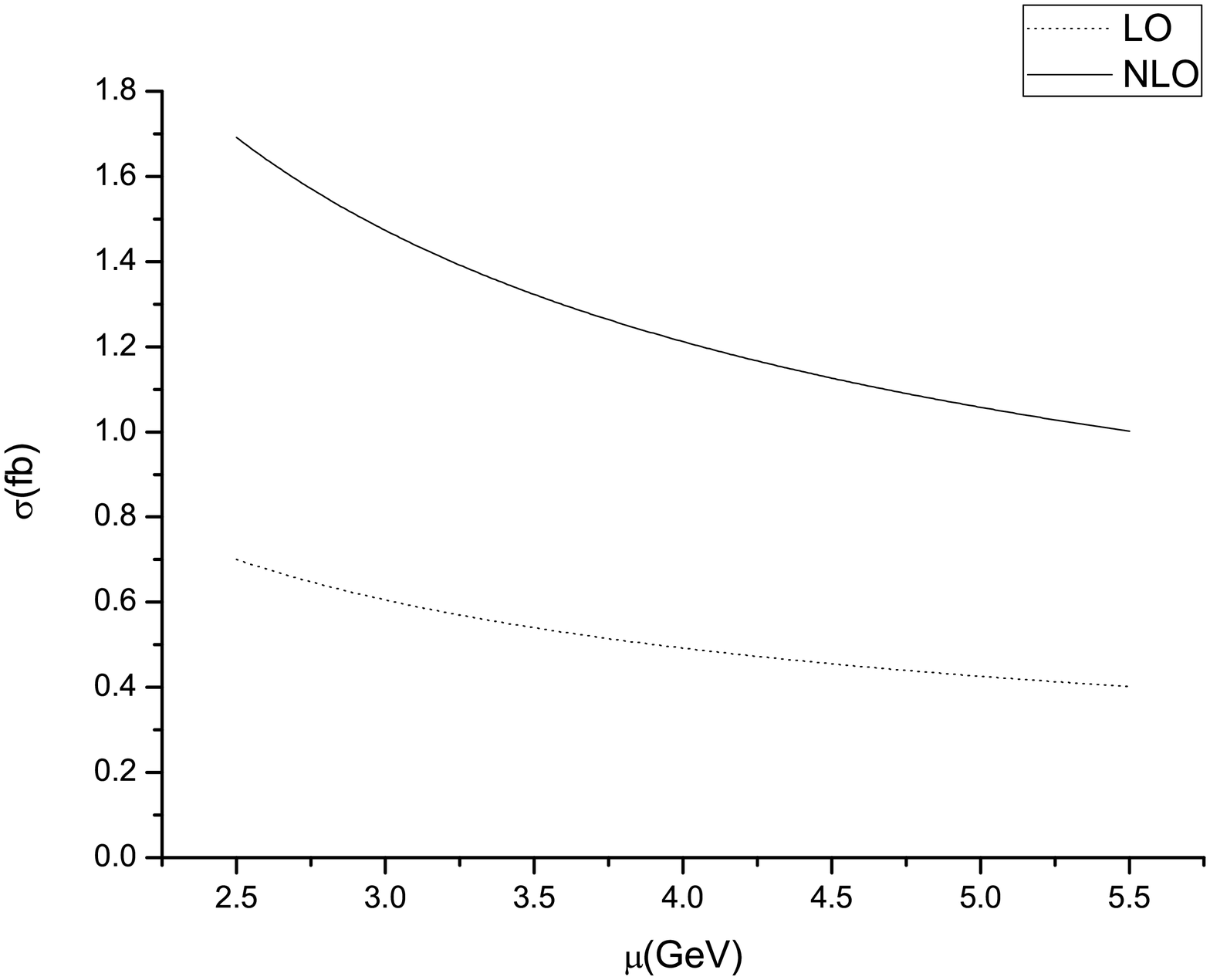}
  \end{minipage}
  }
  \caption{\label{fig:mudep1} The scale $\mu$ dependence of
$e^+e^-\rightarrow \gamma^*\rightarrow \chi_{cJ}(^3P_J^{[1]},^3S_1^{[8]})+ c + \bar{c}+X$ processes at LO and NLO. Here $m_c=1.5\ \textrm{GeV}~\textrm{and}\ \Lambda_{QCD}=296\ \textrm{MeV}$.}
\end{figure*}
\begin{figure*}[hbtp]\centering
\subfloat[$\chi_{c0}(^3S_1^{[8]})+g+g+ X$]{
  \begin{minipage}[t]{.5\linewidth}
    \includegraphics[width=\linewidth]{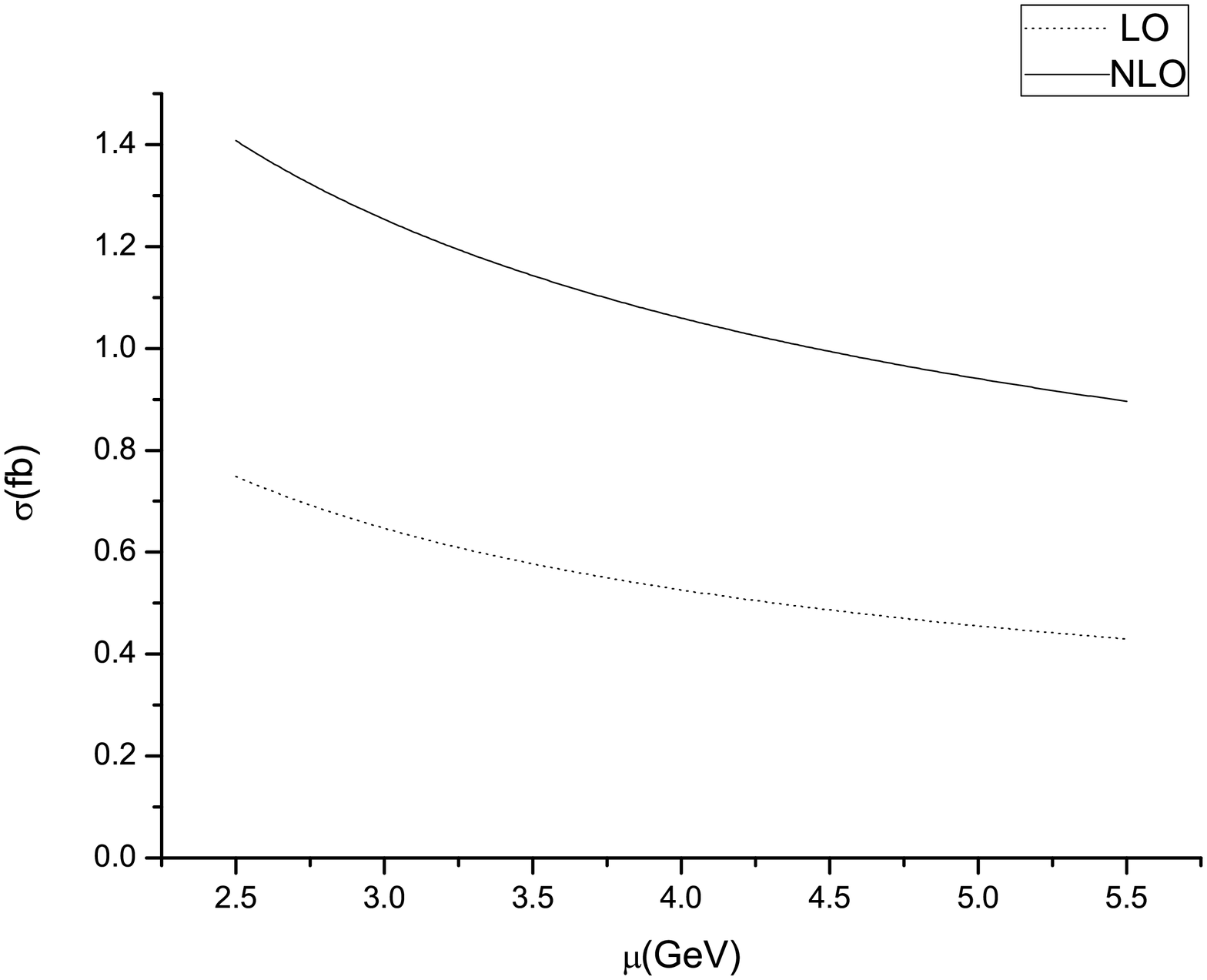}
  \end{minipage}
} \subfloat[$\chi_{c0}(^3S_1^{[8]})+q+\bar{q}+ X$]{
  \begin{minipage}[t]{.5\linewidth}
    \includegraphics[width=\linewidth]{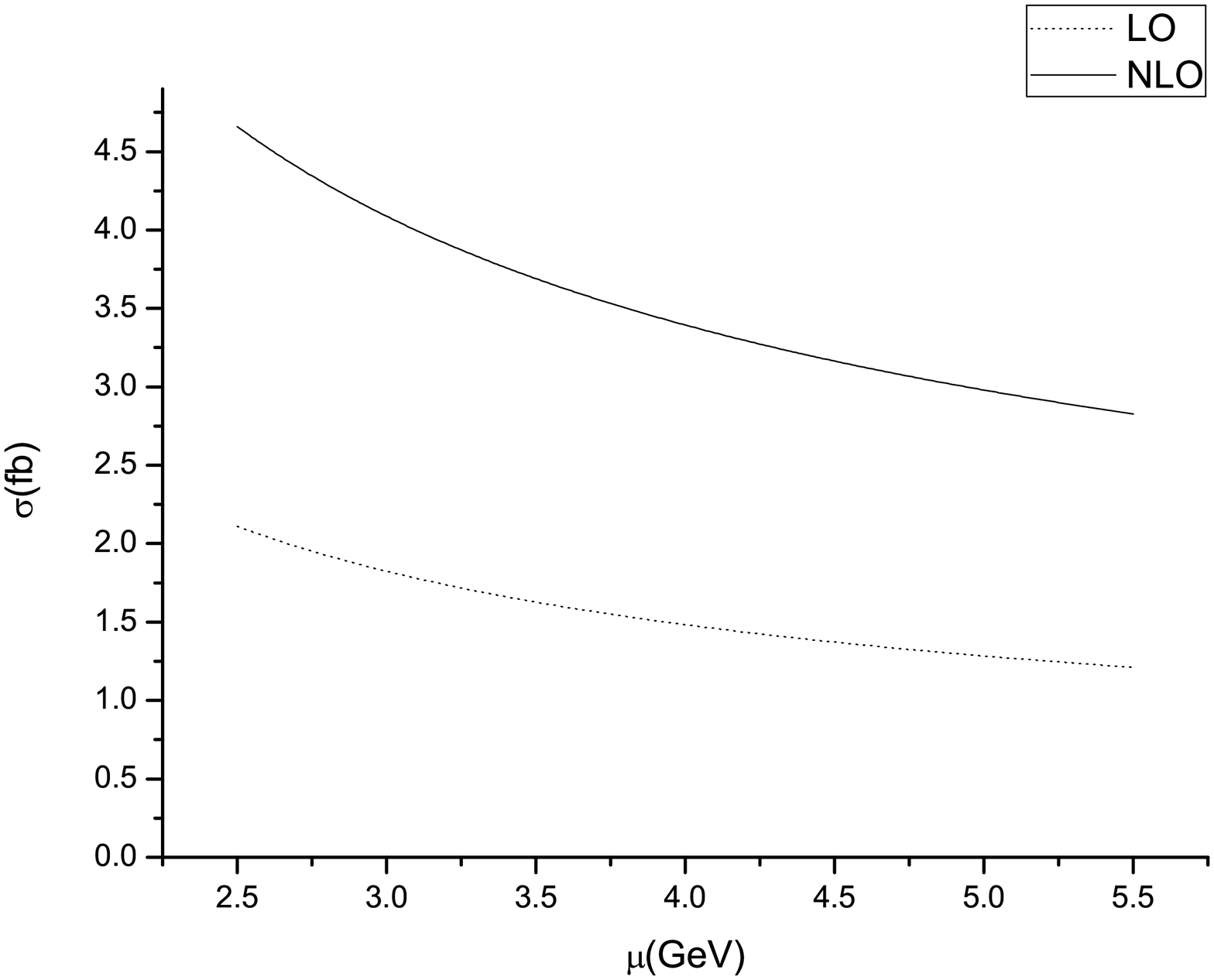}
  \end{minipage}
}
  \caption{\label{fig:mudep2} The scale $\mu$ dependence of
$e^+e^-\rightarrow \gamma^*\rightarrow \chi_{cJ}(^3S_1^{[8]})+ g g(q\bar{q}) + X$ processes at LO and NLO. Here $m_c=1.5\ \textrm{GeV}~\textrm{and}\ \Lambda_{QCD}=296\ \textrm{MeV}$.}
\end{figure*}

For fixed order calculation, one of the main uncertainties in final results come from the scale dependence, which in principle becomes weaker with higher order corrections. In Figures 3-4 we show the renormalization scale $\mu$ dependence of LO and NLO cross sections of $e^+e^-\rightarrow \gamma^*\rightarrow \chi_{cJ}(^3P_J^{[1]},^3S_1^{[8]})+c\bar{c}(g g; q\bar{q}) + X$ inclusive
processes. The Figures exhibit that only the $\mu$ dependence of
$^3P_1^{[1]}$ production process is slightly depressed by NLO QCD
corrections, while no evident improvement for others. This happens, we think, because of the large positive NLO corrections for $^3P_0^{[1]},^3S_1^{[8]}$ and large negative correction for $^3P_2^{[1]}$ production processes, that is the large C in Eq.(\ref{nlo-form}). For these processes, the scale dependence would be reduced when even higher order corrections are taken into account.

In summary, we have calculated the NLO QCD corrections to
$e^+e^-\rightarrow\gamma^*\rightarrow \chi_{cJ}(^3P_J^{[1]},^3S_1^{[8]}) + c\bar{c}(g g; q\bar{q}) + X$ inclusive production processes, especially evaluated the cross sections for $B$ factories in detail. The NRQCD factorization works well, that is all divergences can be properly handled, when both color-singlet and color-octet mechanisms are taken into account. Large positive NLO corrections have been found for $^3P_0^{[1]}$ and $^3S_1^{[8]}$ production processes, while for $^3P_2^{[1]}$ the cross section is substantially depressed by NLO QCD correction. For $\chi_{c0}$ inclusive production, the dominant contribution comes from the color-singlet mechanism, while for $\chi_{c1}$ and $\chi_{c2}$ inclusive production
the dominant contribution comes from the color-octet mechanism. From the results in this work, one can understand why previous BaBar measurement gave only the upper limits for  $\chi_{c1}$ and $\chi_{c2}$ production, and one may also expect that the $\chi_{cJ}$ will be measurable in BELLE II(super-B) experiment.

\newpage
\vspace{0.7cm} {\bf Acknowledgments}

This work was supported in part by National Key Basic Research Program of China under the grant 2015CB856700, and by the National Natural Science Foundation of China(NSFC) under the grants 11175249 and 11375200.

\vspace{1cm}

\end{document}